# Analyzing complex functional brain networks: fusing statistics and network science to understand the brain

**Sean L. Simpson**

*Department of Biostatistical Sciences*
*Wake Forest School of Medicine*
*Winston-Salem, NC*
*and*
*Department of Biostatistics*
*University of North Carolina at Chapel Hill*
*Chapel Hill, NC*
*email:* slsimpso@wakehealth.edu

**F. DuBois Bowman**

*Department of Biostatistics and Bioinformatics*
*The Rollins School of Public Health*
*Emory University*
*Atlanta, GA*
*email:* dbowma3@emory.edu

**Paul J. Laurienti**

*Department of Radiology*
*Wake Forest School of Medicine*
*Winston-Salem, NC*
*email:* plaurien@wakehealth.edu

**Abstract**: Complex functional brain network analyses have exploded over the last decade, gaining traction due to their profound clinical implications. The application of network science (an interdisciplinary offshoot of graph theory) has facilitated these analyses and enabled examining the brain as an integrated system that produces complex behaviors. While the field of statistics has been integral in advancing activation analyses and some connectivity analyses in functional neuroimaging research, it has yet to play a commensurate role in complex network analyses. Fusing novel statistical methods with network-based functional neuroimage analysis will engender powerful analytical tools that will aid in our understanding of normal brain function as well as alterations due to various brain disorders. Here we survey widely used statistical and network science tools for analyzing fMRI network data and discuss the challenges faced in filling some of the remaining methodological gaps. When applied and interpreted correctly, the fusion of network scientific and statistical methods has a chance to revolutionize the understanding of brain function.

## 1. Introduction

As evidenced by the launching of the Human Connectome Project (HCP) by the National Institutes of Health (NIH) in 2009 and the 1000 Functional Connectomes Project in the same year, whole-brain functional magnetic resonance imaging (fMRI) connectivity analyses are key in our understanding of normal brain function as well as alterations due to various brain disorders [1, 2]. fMRI measures localized brain activity by capturing changes in blood flow (hemodynamic response) and oxygenation associated with neural activity. The blood oxygen level-dependent (BOLD) contrast exploits the magnetic properties of oxygenated and deoxygenated blood to capture these changes [3]. The brain is generally parcellated into cubic regions roughly a few millimeters in size called *voxels* in which the brain activity measurements are made across a series of scans. For coarser representations the BOLD signal time series are averaged across voxels within a specified region. Functional connectivity analysis (FC) examines functional associations (e.g., correlations) between time series pairs in specified voxels or regions [4, 5]. Effective connectivity analysis (EC) examines the directed influence of a time series from one region on that from another [5]. Complex functional brain network (or connectivity) analysis is a specific subfield of connectivity analysis in which associations are quantified for all time series pairs to create an interconnected representation of the brain (a brain network). Studying the brain as a network is appealing as it can be viewed as a system with various interacting regions that produce complex behaviors [6, 7]. As with other biological networks, understanding the complex network organization of the brain has profound clinical implications [1, 2, 6, 8].

This emerging area of complex fMRI network analyses has revealed methodological gaps that require the integration of statistical tools with network-based neuroimage analysis. The application of network science to the brain has facilitated our understanding of how the brain is structurally and functionally organized. Furthermore, studying the brain within this framework has already shed light on how some disorders such as Parkinson's disease, schizophrenia, and Alzheimer's disease affect the brain [8-10]. In the case of Alzheimer's disease, the precuneus shows the most reliable changes based on clinical positron emission tomography (PET) imaging [11, 12]. It has been difficult to reconcile this finding with the predominant clinical symptom of memory dysfunction, a cognitive process associated with the hippocampi. However, recent network analyses have discovered that the precuneus is anatomically and physiologically a central hub (highly connected area) in the brain [13]; thus, damage to it can lead to a number of conditions and reverberate throughout many areas of the brain including the hippocampus. In practice, graph metrics such as clustering coefficient, path length and efficiency measures are often used to characterize system properties of brain networks. Centrality metrics such as degree, betweenness, closeness, and eigenvector centrality determine critical areas within the network. Community structure is also essential for understanding network organization and topology.

Network science has led to a paradigm shift in the neuroscientific community, but many statistical issues remain unaddressed [14]. A more rigorous statistical assessment and a greater scientific understanding of how current network models apply to the brain are needed. An integrated appraisal of multiple network metrics should be performed to better understand network structure rather than focusing on univariate assessments. Statistically comparing groups of brain networks while accounting for their complex topologies remains a fertile area for methodological development. In addition to accounting for the dependence structure of networks, a framework in which the effects of multiple variables of interest and local network features (e.g., disease status, age, race, nodal clustering, nodal centrality, etc.) on the overall network structure can be examined concurrently is paramount. In other words, (non)linear modeling and inferential frameworks for brain networks are in their infancy and have yet to be developed to the extent that equivalent tools have been developed for fMRI activation data. The utility of network comparison tools varies by context; thus, outcomes of interest should inform their development. Here we survey widely used statistical and network science tools for analyzing fMRI network data and discuss the challenges faced in filling some of the remaining methodological gaps. These methods necessitate a philosophical shift toward complexity science. In this context, when applied and interpreted correctly, the fusion of network scientific and statistical methods have a chance to revolutionize the understanding of brain function.

For this survey of methods for complex functional brain networks, we delineate network construction methods in Section 2. We then detail descriptive methods for analyzing these constructed networks in Section 3. Modeling and inferential brain network methods are discussed in Section 4. We conclude with a summary discussion including important future directions for complex functional brain network analysis in Section 5.

## 2. Network construction

A brain network is generally represented by an $n \times n$ matrix where $n$ is the number of nodes, with each node corresponding to an area of the brain. The size of the area depends on the chosen parcellation scheme. The entries of the matrix denote the functional similarity between the BOLD fMRI signal time series of all node pairs. A schematic exhibiting how functional brain networks are generated from fMRI time series data is presented in Figure 1. In the following subsections we discuss the basic steps of network construction: 1) defining nodes (brain parcellations), 2) network estimation, and 3) thresholding. We also discuss the use of weighted networks.

### 2.1 Defining nodes

A typical fMRI session measures brain activity in cubic regions a few millimeters in size (voxels) across a series of scans. Appropriate selection and aggregation of these brain regions to represent network nodes remains a methodological challenge [15]. Brain regions are usually selected based on anatomical locations in order to standardize nodal positions across subjects [16]. Careful consideration should be given to the choice of the parcellation scheme (i.e., the scheme used to subdivide the brain into presumably functionally homogeneous regions) which must be consistent across all images [17, 18]. Ideally, a scheme should cover the entire brain with non-spatially overlapping nodes that embody regions with coherent patterns of functional activity. Coarser parcellations select nodes based on regions from an atlas such as the Automated Anatomical Labeling (AAL) atlas [19], while more granular schemes use voxel-based networks [20]. The coarser representations extract the mean time series from each region (averaged across the voxels within that region) conferring computational benefits due to the reduction in dimensionality, and statistical benefits due to the reduction in variability. However, some have suggested that the higher spatial resolution voxel-based networks are more representative of the brain [21]. Others have recommended using an intermediate parcellation scheme in order to balance spatial resolution with the added noise that comes from more resolved networks [15, 22]. [15] constructed intermediate parcellation templates by subdividing AAL atlas based nodes into uniform contiguous micro nodes. Further refinement of voxel-based networks to include more than the typical $n \approx 20,000$ nodes may be of limited scientific value given that cognitive function is dependent on large-scale activation and coactivation of neuronal populations [16].

### 2.2 Network estimation

After selecting a parcellation scheme, the next step in the network construction process is to estimate the network. That is, compute the entries of the $n \times n$ connection matrix which represent the functional relationships between node pairs. The best estimation approach is an area of ongoing research [23, 24]. Methods for estimating functional connectivity between network nodes for complex brain network analysis fall into two categories: association measures and modeling approaches. Linear association measures include *correlation* and *coherence*, while nonlinear measures comprise *mutual information* and *generalized synchronization*. The literature on modeling approaches for quantifying whole-brain functional connectivity remains sparse, though a few important contributions have been made by [25] and others. Prior to network estimation, a band-pass filter is often applied to the time series in order to reduce confounding from physiological noise and subject motion [26]. Here we focus on methods for determining undirected functional connectivity (as distinct from directed or effective connectivity (EC)) as the relatively poor temporal resolution of fMRI data generally precludes accurately inferring causal relations in whole brain networks [27]. We also leave out a discussion of independent component analysis (ICA) and similar methods as their applicability in the context of complex functional brain network analysis is debatable [23, 28]. While ICA provides a useful exploratory technique for examining functional connectivity, it precludes quantitatively assessing differences in overall network structure and investigating how different modules (interconnected clusters of nodes) interact with and transfer information between each other [28].

*Correlation* and *partial correlation* are the simplest and most commonly used association measures to quantify the similarity between the time series or frequency spectra (wavelet correlation) of node pairs [29-31]. Partial correlation is the preferred method as it better distinguishes direct from indirect connections. Conceptually, partial correlation falls in the middle of the association measure-modeling method continuum given that its computation involves time series data from all nodes. It has even been promoted as a surrogate for structural equation modeling (SEM) under light regularization [32]. Computing the inverse of the covariance matrix for the nodal time series provides an efficient way to estimate the partial correlations. If it is expected that the connection matrix is sparse, regularization (e.g., applying the Lasso method) allows differentially shrinking smaller correlation values toward zero. This approach may be useful with short fMRI scanning sessions with shorter nodal time series [23]. While linear partial correlation is sufficient for capturing functional associations in most contexts [30], network metrics such as small-worldness may be biased if careful attention is not paid to constructing appropriate null networks for benchmarking [29, 33].

*Coherence* is the spectral analogue of correlation [34-38], and is defined as

$$C_{t_i t_j} = \frac{|f_{t_i t_j}(\lambda)|^2}{f_{t_i}(\lambda) \cdot f_{t_j}(\lambda)},$$

(1)

where $f_{t_i t_j}(\lambda)$ is the cross-spectrum at frequency $\lambda$ of the time series at nodes $i$ and $j$, and $f_{t_i}(\lambda)$ and $f_{t_j}(\lambda)$ are the respective power spectrums at frequency $\lambda$. This normalized measure takes a value of 0 in the absence of any linear relationship, and 1 if the time series are perfectly related by a linear magnitude and phase transformation. It is generally estimated for either a single frequency range, or multiple ranges with a subsequent combining of results. Use of this measure is more common in EEG and MEG studies where the temporal resolution is much higher.

*Mutual information* (MI) is an often employed nonlinear association measure [39-41] that captures both linear and nonlinear dependencies between nodal time series, and is defined as

$$I(T_i, T_j) = \int_{T_i} \int_{T_j} f(t_i, t_j) \cdot \log_2 \left( \frac{f(t_i, t_j)}{f(t_i) \cdot f(t_j)} \right) dt_i \, dt_j.$$

(2)

Here $f(t_i, t_j)$ is the joint probability density function of the time series at nodes $i$ and $j$, and $f(t_i)$ and $f(t_j)$ are the respective marginal probability density functions. Conceptually, MI quantifies the shared information between time series pairs. That is, it relays the amount of uncertainty remaining about one time series after knowing the other. If no information is shared (i.e., the BOLD signals are independent) then $I(T_i, T_j) = 0$, otherwise $I(T_i, T_j)$ increases as the amount of information shared increases. Normalized MI allows comparing values within and between subjects [40].

*Generalized synchronization* (or state space synchrony) is another nonlinear association measure that quantifies the interdependence between two signals in state space reconstructed mappings [42, 43]. Essentially, each time series is mapped to a set of delay vectors (i.e., the data are resampled at variable time bins) which have the form

$$T_i(b) = (t_i(b), t_i(b+l), t_i(b+2l) \dots, t_i(b + (m-1)l)).$$

(3)

Here $T_i$ denotes the time series from node $i$, $b$ denotes a discrete reference time point, $l$ is the chosen time lag and $m$ is the chosen embedding dimension. Appropriately choosing $l$ and $m$, along with having time series that are generated by a deterministic dynamical system with a smooth attractor, ensures that the delay vectors lie on a smooth manifold in $\mathbb{R}^m$. Synchrony assessment is then based on the level of similarity among the delay vectors (state space reconstructed mappings) for nodal pairs.

Although nonlinear association measures such as synchronization and MI are more sensitive to higher order dependencies than correlation, their practical relevance for fMRI network estimation is debatable and needs futher evaluation [30]. These measures tend to be relatively sensitive to noise and prone to systematic errors such as estimation bias [44, 45]. Moreover, linear approaches often perform well for signals with mild nonlinearity [30, 46].

While *modeling* methods are often employed for directed network estimation (e.g., Bayes Net, multivariate autoregressive, and dynamic causal modeling), they remain relatively limited for undirected functional network estimation. This is likely, in part, due to the general acceptance of the use of association measures in the fMRI brain network literature. However, a few important model-based estimation contributions have been made. [25] developed a modeling approach to improve the estimation of an individual subject's network by leveraging information contained in a group of subjects' time series data. [47] applied Markov models to infer functional connectivity structure. [48] introduced Dynamic Connectivity Regression (DCR) which allows estimating multiple networks across contiguous temporal partitions. Penalized regression methods have also been proposed for network estimation [49].

### 2.3 Thresholding

After estimating a functional brain network from nodal time series, the next step often involves thresholding the connection matrix to remove weak connections and produce an $n \times n$ adjacency matrix $(A_{ij})$ which notes the presence or absence of a functional connection between any two nodes ($i$ and $j$). These binary functional connections are called *edges* or *links* in the network. The sparse binary brain networks resulting from the thresholding process are comprised of strong or "significant" connections and have served as the impetus for many of the network scientific and statistical methods developed thus far [50]. In the case of a dense network generated by eliminating the thresholding step or by employing a very lenient threshold, preserving edge weights is most appropriate (Section 2.4). Weighted versions of the traditional descriptive metrics (described in Section 3) can then be used [14, 50].

Credible network analysis requires careful choice of the thresholding approach as it affects the density of connections and network topology in ways that can yield erroneous conclusions [51]. As with all of the network construction steps, thresholding strategy development is an area of ongoing research. How to assess credibility and determine the "best" strategy remain open questions. The optimal method likely varies with the research question of interest. As a point of clarification, thresholding is sometimes referred to as "network inference" since a network structure is being inferred. This is distinct from the way we use "network inference" in this review which refers to drawing statistical conclusions about an already constructed network or group(s) of networks.

Most thresholding methods fall into three categories: *fixed threshold*, *fixed average degree*, and *fixed edge density* [7, 51]. The *fixed threshold* approach requires selecting a single threshold according to one of three criteria: (1) using a specified significance level (e.g., $\alpha = 0.05$) to retain "significant" connections (with a correction for multiple comparisons); (2) employing a uniform threshold across all networks (e.g., $\hat{\rho} > 0.5$); (3) defining a threshold that minimizes the number of connections while ensuring all nodes are connected to the main component. The major limitation of this fixed threshold approach is that the generated networks generally vary in average degree $k$ (average number of connections) which can confound subsequent comparative analyses [51]. The *fixed average degree* method avoids this problem by varying the threshold applied to each network so that $k$ is fixed across all networks. However, problems arise if the connectivity distributions vary significantly across networks. For example, to maintain the desired average degree for a network dominated by weaker connection strengths with one dominated by stronger connection strengths, the former will have weak, potentially non-significant connections converted to edges while the latter will have strong, significant connections omitted. Alternatively, the *fixed edge density* approach (or *wiring cost*) fixes the proportion of the number of existing edges to the number of possible edges [17]. [21] proposed a thresholding method falling within this category which ensures that $S=\log(n)/\log(k)$ is the same across networks. This relationship is based on the path length of a random network with $n$ nodes and average degree $k$ [52, 53], and can be re-written as $n = k^S$. For networks with the same number of nodes, the methods in this category are equivalent to fixing the average degree.

Thresholding strategy development remains an area of ongoing research given the lack of consensus on the best method. Often researchers will conduct sensitivity analyses to show how their results change over various thresholds and thresholding approaches [20, 31, 51, 54]. Threshold selection based on network size presents one potential solution [55]. However, more work is needed given the sensitivity of network topology to the thresholding process [51].

## 2.4 Weighted networks

Weighted (i.e., continuous) functional brain network analyses involve deriving a weighted adjacency matrix $(W_{ij})$ from the connection matrix. Often the connection matrix itself, or a simple transformation of it, is used, with the edges containing information about connection strengths. These analyses allow avoiding thresholding issues and have gained traction [9, 56, 57] due to recent methodological developments [14, 50, 58]. Such analyses utilize the entire connection matrix rather than a sparse binary adjacency matrix. This option has the benefit of eliminating the thresholding step but poses new challenges. First, the computational burden of the analyses increases considerably since the graph is fully connected. While calculations remain feasible on networks based on brain atlases with a limited number of nodes ($\lesssim 1000$), the computational burden is too great for voxel based networks. Second, the interpretation of analysis results from a fully connected network must be made cautiously. These networks are no longer comparable to sparse networks that depend on connections between clusters for information spread [59]. Given these (and other) computational and methodological challenges that weighted networks pose [7, 50, 58], binary network analysis still dominates the literature. Thresholding weighted networks to remove noise while retaining the continuous "significant" connections may mitigate some of these challenges. However, removing the "weak" links and restricting the range of connection strengths in this manner may limit the power of subsequent analyses and render certain distributional assumptions invalid. Analysis approaches for weighted networks are in their infancy and may prove vital for understanding normal and abnormal brain function.

## 3. Descriptive methods

### 3.1 Functional segregation and integration

Measures of functional segregation and integration are among the most widely used metrics to characterize the topology of fMRI brain networks. Segregation metrics quantify the presence of densely interconnected groups of brain regions, which allow for specialized, segregated neural processing (regional specificity). That is, these measures characterize the brain's local communication ability. *Clustering coefficient* ($C$) [53] and *transitivity* [60] are two such measures based on the number of triangles (three interconnected nodes) in a network. Alternatively, *local efficiency* ($E_{loc}$) [61] provides a scaled analogue of these metrics and is defined as

$$E_{loc} = \frac{1}{n}\sum_{i \in N} E_{loc,i} = \frac{1}{n}\sum_{i \in N} \frac{\sum_{j,h \in N, j \neq i} a_{ij} a_{ih} [d_{jh}(N_i)]^{-1}}{k_i(k_i - 1)}, \tag{4}$$

where $E_{loc,i}$ is the local efficiency of node $i$, $N$ is the set of all nodes in the network, $n$ is the number of nodes, $a_{ij}$ is an indicator function for the existence of an edge between nodes $i$ and $j$, $k_i$ is the degree of node $i$, and $d_{jh}(N_i)$ is the shortest path between nodes $j$ and $h$ that contains only neighbors (connected nodes) of node $i$. $E_{loc}$ ranges from zero to one, with larger values representing more functional segregation.

Functional integration metrics quantify the presence of statistical dependencies between distributed brain regions, indicating the capacity for rapid information transfer (distributive processing). That is, these measures characterize the brain's global communication ability. *Characteristic path length* ($L$), the most commonly used of these measures, is the average shortest distance (minimum number of edges that must be traversed) between all node pairs. *Global efficiency* ($E_{glob}$) [61], a scaled analogue of $L$, is the average inverse shortest distance between node pairs and is defined as

$$E_{glob} = \frac{1}{n}\sum_{i\in N}E_{glob,i} = \frac{1}{n}\sum_{i\in N}\frac{\sum_{j\in N, j\neq i}d_{ij}^{-1}}{n-1},$$ (5)

where $E_{glob,i}$ is the global efficiency of node $i$, $d_{ij}$ is the shortest path between nodes $i$ and $j$, and $N$ and $n$ are defined as before. Like $E_{loc}$, $E_{glob}$ also ranges from zero to one, with larger values representing more functional integration. Weighted and directed analogues for $C$ [62-64], transitivity, $E_{loc}$, $L$, and $E_{glob}$ have also been developed [50].

### 3.2 Small-worldness

The brain is thought to optimize information transfer by maximizing functional segregation and integration while minimizing wiring cost [65]. Small-worldness is often the term used to describe such a design that enables distributive processing and regional specificity. The seminal paper by [53] introducing this small-world idea catalyzed the use of network science in many disciplines including neuroscience. Subsequently, [67] introduced the small-world measure, $\sigma$, to quantify this property. Conceptually, a small-world network is one that is more clustered than a random network while still having approximately the same characteristic path length as a corresponding random network. Mathematically it is defined as

$$\sigma = \frac{C/C_{\text{rand}}}{L/L_{\text{rand}}},$$ (6)

where $C$ and $C_{\text{rand}}$ are the clustering coefficients, and $L$ and $L_{\text{rand}}$ are the characteristic path lengths of the respective network of interest and random (null) network used for benchmarking. Random (null) networks are commonly generated such that they have the same degree distribution as the original network in order to avoid confounding network structure [66]. Though, appropriately constructing these random networks depends, in part, on the method used for estimating the original network [29]. Arbitrary thresholds are often set to distinguish values of $\sigma$ (usually $\sigma > 1$) deemed to signify small-worldness. However, work by [67] provides steps towards formally quantifying and testing for the small-world property.

Lattice networks are also used as null networks [68] given that they provide a better benchmark for assessing network clustering. For this reason, [69] developed an alternative small-world measure defined as

$$\omega = \frac{L_{\text{rand}}}{L} - \frac{C}{C_{\text{latt}}},$$ (7)

where the clustering coefficient of the original network ($C$) is now indexed against a corresponding lattice network $(C_{\text{latt}})$. This scaled metric, $\omega \in [-1, 1]$, takes values close to zero for small-world networks, positive values for more random networks, and negative values for more regular, or lattice-like, networks.

### 3.3 Resilience measures

Infrastructural properties of functional brain networks determine the capacity of localized brain injury or degeneration to affect overall brain capabilities. Complex network analysis allows characterizing these properties with several topological measures that assess network vulnerability to insult. The *degree distribution* [70] is the most commonly assessed property. Empirically, these distributions follow a power-law (Pareto distribution) or exponentially truncated power-law in fMRI brain networks [21]. Networks with a power-law degree distribution tend to be resilient to random injury (i.e., random removal of nodes), but vulnerable to injuries that target network hubs (highly connected nodes). Those with an exponentially truncated power-law distribution maintain resilience to random injuries, while being slightly less vulnerable to injuries that target network hubs given that the hubs tend to be less connected than in the power-law counterpart. Despite the slightly disparate implications of the two distributions for network robustness, rigorous statistical assessment of distributional goodness-of-fit (GOF) in the literature is sparse. Recent work by [71, 72] formally quantifying and testing for GOF will likely change this trend.

The *assortativity coefficient* $(R_{ij})$ is another widely used measure of network resilience [60, 73]. It quantifies the correlation between the degrees of connected nodes and is defined as

$$R_{ij} = \frac{l^{-1}\sum_{(i,j)\in L}k_i k_j - \left[l^{-1}\sum_{(i,j)\in L}\frac{1}{2}(k_i + k_j)\right]^2}{l^{-1}\sum_{(i,j)\in L}\frac{1}{2}(k_i^2 + k_j^2) - \left[l^{-1}\sum_{(i,j)\in L}\frac{1}{2}(k_i + k_j)\right]^2}, \tag{8}$$

where $l$ is the number of edges (links), $L$ is the set of all edges in the network, and $k_i$ and $k_j$ are the degrees of nodes $i$ and $j$ respectively. Assortative networks (those with $R_{ij}$ positive and closer to 1) imply the existence of a resilient core of interconnected high-degree hubs. Conversely, disassortative networks (those with $R_{ij}$ negative and closer to $-1$) imply the existence of more vulnerable high-degree hubs due to their wider distribution throughout the network. Weighted and directed extensions of $R_{ij}$ are discussed in [74] and [73] respectively.

### 3.4 Graph centrality and information flow

Centrality measures quantify the relative importance of a given node in a brain network for the transfer of information. For fMRI networks (and biological networks more generally), four classical centrality assessment metrics are used: degree centrality, betweenness centrality [75], closeness centrality [76], and eigenvector centrality [77]. These measures contain numerous extensions which fall into two main categories (Table 1): radial and medial measures [78]. Radial measures quantify potential information transfer originating from or terminating at a given node, whereas medial measures quantify potential information transfer through a given node. Radial measures comprise degree, closeness, and eigenvector centrality, while medial measures include betweenness centrality metrics. Brain network studies frequently employ these centrality measures due to their implications in variety of diseases [79-82].

Proper metric choice depends on the type of information transfer that a system supports: *serial transfer*, *serial duplication*, or *parallel duplication* (see Figure 2) [83]. *Serial transfer* refers to an information flow pattern in which a node transfers information to only one connected node at time (e.g., package delivery). *Serial duplication* has this same one-to-one information exchange, but the information also remains at the source node (e.g., transmission of a virus). Information is also replicated in *parallel duplication*, though it spreads to all connected nodes (e.g., an email broadcast). It is our contention that neuronal physiology makes it likely that the brain uses parallel duplication to transmit information [7, 84, 85].

### 3.5 Community structure

Functional brain networks subdivide into interconnected communities (modules) that allow an efficient division of labor [86, 87]. This community structure arises from network partitions that maximize the number of within-community nodal connections while minimizing the number of between-community connections. Detecting community structure presents a daunting task given that it is a non-deterministic polynomial-time (NP-hard) problem [88], and that the number and size of communities are unknown. However, many optimization algorithms have proven useful [89]. The Girvan-Newman method delineates communities based on the *edge betweenness* of nodes [86]. This algorithm provides reasonable accuracy, but is limited to smaller networks due to its computational intensiveness. Modularity maximization, one of the most widely used methods in the brain network literature, determines community structure by optimizing the *modularity* statistic (illustrated in Figure 3)

$$Q = \sum_{u \in M} \left[ e_{uu} - \left( \sum_{v \in M} e_{uv} \right)^2 \right],$$
(9)

where $M$ indexes the set of nonoverlapping modules from the fully subdivided network, and $e_{uv}$ denotes the proportion of all edges that connect nodes in module $u$ with those in module $v$ [88, 90]. While popular, this approach is limited in its ability to detect relatively small communities [91], and encounters degeneracy issues for partitions with high modularity [92]. [93] introduced $Q_{cut}$ to address the former (resolution limit) issue. Their approach combines spectral graph partitioning and local search methods to optimize $Q$. Weighted and directed analogues of $Q$ have also been developed [94, 95]. Surprise ($S$) maximization [96] and the Louvain method [97] are two other community detection optimization algorithms that perform well across a wide range of applications. This list of algorithms for nonoverlapping community detection is not exhaustive, but representative of those used in the functional brain network literature. Given that the validity of detection algorithms varies with network structure, it is unclear which is most appropriate for functional brain network data.

In reality, many brain areas likely belong to multiple modules simultaneously given that they can perform several roles [98]. Acknowledging the occurrence of this phenomenon in networks across a wide variety of areas, [99] developed a quickly adopted algorithm for detecting overlapping modular network structure which employs the clique percolation method (CPM). Alternatively, the Order Statistics Local Optimization Method (OSLOM) identifies (potentially) overlapping communities based on the relative probability that a node connects to a given network substructure compared with this connection likelihood in a comparable random network [100]. ModuLand provides a conceptually different approach to overlapping community structure that groups nodes into modules based on their mutual influence (highly interconnected nodes are considered mutually influential) [101].

Assessing the consistency of community structure within or across subjects presents a challenge that requires innovative approaches. The approximation algorithms employed to detect community structure can produce different results across multiple runs for the same subject [102]. Inter-subject variability makes across-subject analyses even more difficult as the number, size, and composition of modules may vary widely. [103] offered an approach to understand dynamic change in community structure that quantifies nodal stability within a community over time or across multiple realizations. [102] developed an alternative approach, called scaled inclusivity, to assess community structure consistency within and across subjects. While both approaches have proven useful [28, 104], more work in this area is needed.

Although identifying modular structure provides information about labor division within a network, assessing nodal roles within their given communities allows for even more resolved insight. Functional cartography is a classification scheme that determines nodal roles based on their connectivity patterns [105, 106]. Each node is labeled as one of seven types: (R1) ultra-peripheral nodes, (R2) peripheral nodes, (R3) nonhub connecter nodes, (R4) nonhub kinless nodes, (R5) provincial hubs, (R6) connector hubs, and (R7) kinless hubs. Further details regarding the classification procedure are provided in [7, 105, 106].

*3.6 Metrics as random variables*

When characterizing functional brain networks with descriptive metrics like those discussed in this section, the fact that the underlying network is estimated (see Section 2.2) is largely ignored. That is, the fact that these metrics are functions of an estimated network and thus are estimates themselves with certain probability distributions is not taken into account. Other than [107], to our knowledge, no work has been done on propagating the estimation error from the network to the network metrics. More formally, we denote the true network as $Y = (N, E)$ and the estimated network as $\widehat{Y} = (N, \widehat{E})$, where $N$ is the set of all nodes and $E$ the set of all edges. Properly defining the true network requires setting a priori the definition of nodes $N$, and the method(s) employed to define the edges $E$. That is, we want to propagate conditional error given these choices. The descriptive metrics are functions of the estimated network which we denote by $g(\widehat{Y})$, with $g(Y)$ representing the true value of the network metric. It is then of interest to examine several properties of the quantity $\Delta = g(\widehat{Y}) - g(Y)$: (1) Does $E(\Delta) = 0$? (2) What is the distribution of $\Delta$? (3) What are the confidence intervals for $g(Y)$? In other words, propagating error appropriately allows making formal inferential decisions about network values and gaining a better understanding of topological variability. [107] quantified the propagation of error from network estimation to the density metric (proportion of the number of existing edges to the number of possible edges) for a correlation-based network. Even deriving the distribution of this simple descriptive metric poses a difficult statistical challenge. Deriving the distributions of the more complicated metrics discussed here will prove daunting. This area of network error propagation has barely been tapped and provides extremely fertile ground for statistical research.

## 4. Modeling and inferential methods

The emerging area of complex functional brain network analysis has created modeling and inferential gaps that require the integration of statistical tools with network-based neuroimaging analysis. As observed by [58], "a statistically principled way of conducting brain network analysis is still lacking." Also, as noted by [14], "between-subject comparisons in studies of brain networks will require the development of accurate statistical tools." To date, the amount of statistical work done in these areas has not been commensurate with their level of importance [51]. Most current approaches to modeling and comparing functional brain networks either rely on a specific extracted summary metric [17, 79, 108, 109] which may lack clinical use due to low sensitivity and specificity, or on mass-univariate edge-based comparisons that ignore the inherent topological properties of the network while also yielding little power to determine significance [110]. While some univariate approaches have proven useful [111], gleaning deeper insights into normal and abnormal changes in complex brain function demands methods that match the complexity of the data while allowing for tractable results. Fusing multivariate statistical approaches with network science presents the best path to develop these methods. In the following subsections we survey the univariate, multivariate, and doubly multivariate (longitudinal networks/network dynamics) tools available for analyzing fMRI network data noting gaps where they exist. We also discuss potential approaches to fill these gaps with the development of new methods and modification of existing methods from other scientific areas. As noted earlier, "network inference" is an ambiguous term that can refer to network construction (Section 2). Here we use the term to refer to drawing conclusions about already constructed networks or group(s) of networks.

## 4.1 Univariate methods

As previously noted, most modeling and inferential methods employed in the analysis of functional brain networks are univariate in nature. Often descriptive metrics at the network or nodal level (like those discussed in Section 3) are compared across groups using ANOVA like techniques [112] or the estimated connectivity values (detailed in Section 2) themselves are compared at each edge with a multiple testing correction applied [110]. The network-based statistic (NBS) and spatial pairwise clustering (SPC) methods afford more power than a traditional edge-based approach by looking for sub-network differences [111] (see Figure 4 for a graphic example). Conceptually, they are network analogues of cluster-based thresholding of statistical parametric maps (a mass-univariate testing procedure for brain activation). The NBS and SPC both follow a similar procedure: 1) admit edges with a test statistic surpassing a set threshold to a set of supra-threshold connections; 2) search for distinct clusters in the set of supra-threshold connections; 3) compute family-wise error corrected p-values for each cluster via permutation testing. In other words, both approaches test for an experimental effect at the cluster level as opposed to the edge level. Differences in the methods lie in the criteria used to define clusters. [111] further describes and compares the two approaches. The NBS generally better identifies experimental effects spanning multiple interconnected regions; while the SPC more accurately discerns effects between isolated region pairs. Figure 5 provides an illustrative comparison of the two approaches. Both are gaining traction in the brain network literature [113-115].

## 4.2 Multivariate methods

While univariate approaches like the NBS and SPC are useful and provide a foundation for future developments, further elucidation of normal and abnormal changes in complex brain function requires more sophisticated multivariate methods. Massively univariate analyses do not allow harnessing the full power of brain networks which lies in understanding their complex organization. The complex topological properties of the system confer much of its functional ability. For example, functional connections may be lost due to an adverse health condition but compensatory connections may develop as a result in order to maintain topological consistency and functional performance. Ultimately, a multivariate explanatory and predictive (non)linear modeling framework is needed that accounts for the complex dependence structure of networks and allows assessing the effects of multiple variables of interest and local network features (e.g., demographics, disease status, nodal clustering, nodal centrality, etc.) on the overall network structure. That is, if we have

$$\text{Data} \begin{cases} \boldsymbol{Y}_i : \text{network of subject } i \\ \boldsymbol{X}_i : \text{covariate information (metrics, demographics, etc.)} \end{cases},$$

we want the ability to model the probability density function of the network given the covariates $P(\boldsymbol{Y}_i | \boldsymbol{X}_i, \boldsymbol{\theta}_i)$, where $\boldsymbol{\theta}_i$ are the parameters that relate the covariates to the network structure. Optimal methods within this framework likely vary by context; thus, outcomes of interest should inform their development.

Exponential random graph models (ERGMs) provide one such multivariate approach to modeling functional brain networks [116]. They have the form of the well-studied regular exponential family given below:

$$P_{\boldsymbol{\theta}}(\boldsymbol{Y} = \boldsymbol{y} | \boldsymbol{X}) = \kappa(\boldsymbol{\theta})^{-1} \exp\{\boldsymbol{\theta}^{\mathrm{T}} \mathbf{g}(\boldsymbol{y}, \boldsymbol{X})\}. \tag{10}$$

Here $\boldsymbol{Y}$ is an $n \times n$ ($n$ nodes) random symmetric adjacency matrix representing a brain network from a particular class of networks, with $Y_{ij} = 1$ if an edge exists between nodes $i$ and $j$ and $Y_{ij} = 0$ otherwise. The probability mass function (pmf) $\left(P_{\boldsymbol{\theta}}(\boldsymbol{Y} = \boldsymbol{y} | \boldsymbol{X})\right)$ of this class of networks is a function of the prespecified network features defined by $\mathbf{g}(\boldsymbol{y}, \boldsymbol{X})$. This vector of prespecified explanatory metrics can consist of covariates that are functions of the network $\boldsymbol{y}$ (e.g., number of paths of length two) and nodal covariates ($\boldsymbol{X}$) (e.g., brain location of the node). The parameter vector $\boldsymbol{\theta}$, associated with $\mathbf{g}(\boldsymbol{y}, \boldsymbol{X})$, quantifies the relative significance of the network features in explaining the structure of the network after accounting for the contribution of all other network features in the model. More specifically, $\theta$ indicates the change in the log odds of an edge existing for each unit increase in the corresponding explanatory metric. If the $\theta$ value corresponding to a given metric is large and positive, then that metric plays a considerable role in explaining the network architecture and is more prevalent than in the null model (random network with the probability of an edge existing ($p$) = 0.5). Conversely, if the $\theta$ value is large and negative, then that metric still plays a considerable role in explaining the network architecture but is less prevalent than in the null model. Consequently, inferences can be made about whether certain local features/substructures are observed in the network more than would be expected by chance [117], enabling hypothesis development regarding the biological processes that produce these structural properties. The normalizing constant $\kappa(\boldsymbol{\theta})$ ensures that the probabilities sum to one. This approach allows representing the global network structure by locally specified explanatory metrics, thus providing a means to examine the nature of networks that are likely to emerge from these effects.

[116] illustrated the utility of ERGMs for modeling, analyzing, and simulating functional brain networks. [118] showed how ERGMs can be used to produce group-based "representative" networks that capture important average topological properties and nodal distributions of those properties in a group of networks better than the standard approaches. Simply averaging the connectivity matrices, and thresholding the resulting matrix, fails to produce an accurate "summary" network [56 (Appendix A), 118]. The need for these representative networks is well documented [31, 112, 119-124]. Despite the utility of the ERGM framework for efficiently representing complex network data and inherently accounting for higher order dependence/topological properties, it has several limitations within the brain network context. Multiple-subject comparisons can pose problems given that these models were originally developed for the modeling of one network at a time [116]. Additionally, the amount of programming work increases linearly with the number of subjects since ERGMs must be fitted and assessed for each subject individually [118]. Incorporating novel metrics (perhaps more rooted in brain biology) may be difficult due to degeneracy issues that may arise [125, 126]. While well-suited for substructural assessments, edge-level examinations remain difficult with these models. Moreover, most ERGM developments have been for binary networks; approaches for weighted networks have been proposed but remain in their infancy [127, 128].

While attempting to address the ERGM limitations directly is important, a mixed modeling approach may provide a more flexible, complementary method [129]. As with ERGMs, mixed modeling approaches for network data have mostly been developed for analyzing an individual network [129-131]. Adapting this framework to our multi-network context presents challenges, but addresses many of the ERGM drawbacks: mixed models are well-suited for edge-level examinations and multiple subject comparisons; novel metrics can be easily incorporated; and they are easily adaptable to weighted and longitudinal networks. However, unlike ERGMs, they do not inherently account for the higher order dependence/topological properties of networks. [130] and others have begun addressing this issue, but more work is needed. We are currently working on adapting mixed modeling network approaches to the analysis of fMRI whole-brain network data. Initially, we are assuming that partial correlations are employed to estimate the network with negative correlations set to 0 (no connection) as it is standard to examine negatively correlated networks separately due to their differing topological properties [7]. Given that we have positively weighted networks, with negative weights set to 0/no connection, we will be proposing a two-part mixed-effects model in order to model both the probability of a connection (presence/absence) and the strength of a connection if it exists. These models enable quantifying the relationship between an outcome (e.g., disease status) and the functional connectivity between brain areas while reducing spurious correlations through inclusion of confounding covariates. They also enable prediction about an outcome based on connectivity structure and vice versa. Several two-part models have been proposed in the literature for a variety of applications [132, 133]. However, they have yet to be developed for networks in general or, more specifically, for functional brain networks.

Let $Y_{ijk}$ represent the *strength* of the connection (quantified as the partial correlation in our case) and $V_{ijk}$ indicate whether a connection is present (*presence* variable) between node $j$ and node $k$ for the $i^{\text{th}}$ subject. Thus, $V_{ijk} = 0$ if $Y_{ijk} = 0$ (or $Y_{ijk} \leq c$ if thresholding), and $V_{ijk} = 1$ if $Y_{ijk} > 0$ (or $Y_{ijk} > c$ if thresholding) with conditional probabilities

$$P(V_{ijk} = v_{ijk}|\boldsymbol{\beta}_v; \boldsymbol{d}_{vi}) = \begin{cases} 1 - p_{ijk}(\boldsymbol{\beta}_v; \boldsymbol{d}_{vi}) & \text{if } v_{ijk} = 0 \\ p_{ijk}(\boldsymbol{\beta}_v; \boldsymbol{d}_{vi}) & \text{if } v_{ijk} = 1, \end{cases} \tag{11}$$

where $\boldsymbol{\beta}_v$ is a vector of population parameters (fixed effects) that relate the probability of a connection to a set of covariates ($\boldsymbol{X}_{ijk}$) for each subject and nodal pair (dyad), and $\boldsymbol{d}_{vi}$ is a vector of subject- and dyad-specific parameters (random effects) that capture how this relationship varies about the population average ($\boldsymbol{\beta}_v$) by subject and dyad. Hence, $p_{ijk}(\boldsymbol{\beta}_v; \boldsymbol{d}_{vi})$ is the probability of a connection between nodes $j$ and $k$ for subject $i$. We then have the following logistic mixed model (part I model) for the probability of this connection:

$$logit(p_{ijk}(\boldsymbol{\beta}_v; \boldsymbol{d}_{vi})) = \boldsymbol{X}'_{ijk}\boldsymbol{\beta}_v + \boldsymbol{1}'\boldsymbol{d}_{vi}. \tag{12}$$

For the part II model, which aims to model the strength of a connection given that there is one, we let $S_{ijk} = [Y_{ijk}|V_{ijk} = 1]$. In our case, the $S_{ijk}$ are the values of the partial correlation coefficients between nodes $j$ and $k$ for subject $i$. We can then use Fisher's Z-transform, denoted as $FZT$, and assume normality (we have empirically observed normality in strength distributions) for the following mixed model (part II model)

$$FZT(S_{ijk}(\boldsymbol{\beta}_s; \boldsymbol{d}_{si})) = \boldsymbol{X}'_{ijk}\boldsymbol{\beta}_s + \boldsymbol{1}'\boldsymbol{d}_{si} + e_{ijk}, \tag{13}$$

where $\boldsymbol{\beta}_s$ is a vector of population parameters that relate the strength of a connection to the same set of covariates ($\boldsymbol{X}_{ijk}$) for each subject and nodal pair (dyad), $\boldsymbol{d}_{si}$ is a vector of subject- and dyad-specific parameters that capture how this relationship varies about the population average ($\boldsymbol{\beta}_s$) by subject and dyad, and $e_{ijk}$ accounts for the random noise in the connection strength of nodes $j$ and $k$ for subject $i$. We assume that $\boldsymbol{d}_{vi}$, $\boldsymbol{d}_{si}$, and $\boldsymbol{e}_i = \{e_{ijk}\}$ are normally distributed. Specifically, $\boldsymbol{e}_i \sim N(0, \boldsymbol{\Sigma} = \sigma^2 \boldsymbol{\Omega})$ (here $\sigma$ represents a standard deviation and not the small-world metric), where the matrix $\boldsymbol{\Omega}$ contains the correlations between connection strengths of dyads and may be modeled with the *linear exponent autoregressive* (LEAR) structure that we have found to work well for correlated neuroimaging data [134, 135].

The covariates ($\boldsymbol{X}_{ijk}$) used to explain and predict both the presence and strength of connection generally fall into four categories: 1) $Net$: network measures (Section 3); 2) $COI$: Covariate of Interest (e.g., disease status); 3) $Int$: Interactions of the Covariate of Interest with the metrics in 1); and 4) $Con$: Confounders (race/ethnicity, gender, etc.). For the random-effects vectors we have that $\boldsymbol{d}_{vi} = [\boldsymbol{d}_{vi,net}\ \boldsymbol{\delta}_{vi,j}\ \boldsymbol{\delta}_{vi,k}\ \boldsymbol{\phi}_{vi,jk}]'$ and $\boldsymbol{d}_{si} = [\boldsymbol{d}_{si,net}\ \boldsymbol{\delta}_{si,j}\ \boldsymbol{\delta}_{si,k}\ \boldsymbol{\phi}_{si,jk}]'$. Here $\boldsymbol{d}_{vi,net}$ and $\boldsymbol{d}_{si,net}$ contain the subject-specific parameters that capture how much the relationships between the network measures in 1) and the *presence* and *strength* of a connection vary about the population relationships ($\boldsymbol{\beta}_v$ and $\boldsymbol{\beta}_s$) respectively. We let $\boldsymbol{\delta}_{vi,j}$ and $\boldsymbol{\delta}_{si,j}$ contain nodal-specific parameters that represent the propensity for node $j$ (of the given dyad) to be connected and the magnitude of its connections respectively, and $\boldsymbol{\delta}_{vi,k}$ and $\boldsymbol{\delta}_{si,k}$ contain nodal-specific parameters that represent the propensity for node $k$ (of the given dyad) to be connected and the magnitude of its connections respectively. Finally, $\boldsymbol{\phi}_{vi,jk}$ and $\boldsymbol{\phi}_{si,jk}$ contain dyad-specific parameters that account for higher-order dependence/topological properties inherent in network data in general [130] and particularly in brain networks [6]. Additional covariates can easily be incorporated as guided by the biological context.

One of the main challenges in this approach is properly specifying $\boldsymbol{\phi}_{vi,jk}$ and $\boldsymbol{\phi}_{si,jk}$ in order to accurately account for the topological properties inherent in brain networks. Extant candidates include the bilinear effect $\left(\boldsymbol{\phi}_{i,jk} = \boldsymbol{z}_j' \boldsymbol{z}_k\right)$; distance model $\left(\boldsymbol{\phi}_{i,jk} = -\,|\boldsymbol{z}_j - \boldsymbol{z}_k|\right)$; and projection model $\left(\boldsymbol{\phi}_{i,jk} = \boldsymbol{z}_j' \boldsymbol{z}_k / |\boldsymbol{z}_k|\right)$. Conceptually, these candidates attempt to capture third-order dependence patterns, such as clustering, present in network data. See [130, 136, 137] for further details on the respective constructions. As other potential candidates arise, they will also be examined. While development of this mixed modeling framework remains nascent, it will fill a critical gap in the fMRI brain network literature and serve as a foundation for future methodological work in the area.

Graph subspace approaches that fall outside a general explanatory and predictive modeling framework may also prove useful for comparing groups of functional brain networks. [138] generalized Kronecker product graph models (KPGMs) to better capture the natural variability across a population of networks. However, this approach has yet to be ported into the brain network analysis domain. [139] proposed representing fMRI brain networks as self-organizing maps (SOMs) and then comparing these maps within a Frechean inferential framework. They compute a mean SOM in each group as a Frechet mean with respect to a metric on the space of SOMs and then compare groups by permuting group labels and using different distance functions to quantify map differences. That is, they conduct a single test to identify whether brain network regions are different at a multivariate level by comparing two non-parametric unsupervised representations of the original data. We are currently working on a similar approach in which Jaccard index values are compared using a permutation of the group representations. Conceptually describing and interpreting multivariate results often present a challenge in general and particularly in the brain network context. These challenges include (but are not limited to) difficulties in simultaneously interpreting several interrelated outcome network variables or complex combinations of these variables, assessing the robustness of multivariate models (diagnostics are less straightforward than in univariate settings), disentangling macro (network-level) and micro (edge or node level) results, and drawing simultaneous inference about topological and spatial differences in whole-brain networks. However, the elucidative benefits of multivariate approaches warrant the additional effort.

### *4.3 Doubly multivariate methods*

The connectivity structure of functional brain networks changes across time and different task conditions. Evaluating how such dynamic changes in network connectivity relate to brain dysfunction will provide insight into underlying mechanisms. The methods discussed in Section 4.2 allow multivariately modeling a static brain network, but extensions are needed to address the dynamic/longitudinal component present in many contexts. These approaches are considered doubly multivariate since there exists multivariate dependencies within a network **and** across networks over time. Other, more specific terms for these types of data include longitudinal network, network-temporal, network dynamics, and network panel data. The term "network dynamics" is ambiguous as it can refer to dynamics *of* a network (our case here) or dynamics *on* a network (i.e., information flow; Section 3.4). Henceforth we will refer to the analysis of functional brain networks across time/task as longitudinal network analysis.

  While advances in longitudinal network analysis have been made in other areas [140-143], to our knowledge, no methods have been developed for brain networks. As noted in Section 2.2, [48] introduced a method which allows estimating connectivity among specified ROIs within contiguous temporal partitions; however, their approach is not intended to model and draw inference from fully constructed longitudinal whole-brain networks. One potential approach to analyzing longitudinal brain networks might be to construct a network of networks as done in [103] and then apply one of the techniques from Section 4.2. Other potential methods may include adaptations of the approaches from other areas as was done for ERGMs [116, 118] and mixed models (Section 4.2) in the static case. The stochastic actor-oriented models for social network dynamics proposed by [140, 141] provide one such potentially adaptable method. Temporal ERGMs (TERGMs), a temporal extension of ERGMs, provide another, similarly adaptable method [142, 144]. TERGMs have the following form:

$$P_{\boldsymbol{\theta}}\big(\boldsymbol{Y}^t|\boldsymbol{Y}^{t-k},\ldots,\boldsymbol{Y}^{t-1}\big) = \kappa\big(\boldsymbol{\theta},\boldsymbol{Y}^{t-k},\ldots,\boldsymbol{Y}^{t-1}\big)^{-1}\exp\big\{\boldsymbol{\theta}^{\mathrm{T}}\mathbf{g}\big(\boldsymbol{y}^t,\boldsymbol{y}^{t-1},\ldots,\boldsymbol{y}^{t-k}\big)\big\}. \quad (14)$$

Here $P_{\boldsymbol{\theta}}\big(\boldsymbol{Y}^t|\boldsymbol{Y}^{t-k},\ldots,\boldsymbol{Y}^{t-1}\big)$ is the probability mass function for the network at a given time point $t$, conditioned on the previous $k$ realizations. Thus, it is assumed that the network at time $t$ is independent of networks more than $k$ time periods away given knowledge of the networks at time points $t-k$ through $t-1$. That is, the information contained in networks prior to time point $t-k$ is assumed to be contained in the intermediate networks of time points $t-k$ through $t-1$. In this manner, time dependence in an individual network can be modeled and accounted for. While useful in many contexts, the actor-oriented and TERGM approaches suffer from many of the same ERGM drawbacks discussed in the previous section. Contrastingly, [143] proposed a more statistical mixed modeling approach in which the model parameters represent expectations and covariances of edge measurements. While their complementary method confers a more flexible inferential and modeling framework (as discussed for mixed models in the previous section), it does not account for the higher-order dependencies inherent in networks. Additionally, as with the actor-oriented and TERGM approaches, the main adaptive difficulty lies in extending the framework to the multi-network context of brain network analysis.

Extending the mixed model framework delineated in the previous section provides another potential approach to the analysis of longitudinal complex brain network data. For example, elucidating task-based dynamic changes in network connectivity that relate to brain dysfunction requires methods that can model network structure variability within and across tasks. Both sources of variation can be simultaneously modeled with a Kronecker product covariance structure [135,145-147]. Following the notation in equations 13 and 14, we assume that the random error vector $\boldsymbol{e}_i \sim N(\boldsymbol{0}, \boldsymbol{\Sigma} = \sigma^2(task)[\boldsymbol{\Gamma} \otimes \boldsymbol{\Omega}])$, where $\sigma^2(task)$ is the variability of connections strengths across the network dyads and varies by task, $\boldsymbol{\Gamma}$ contains the correlations between tasks, and $\boldsymbol{\Omega}$ again contains the correlations between connection strengths of dyads. We adopt this technique of modeling the correlation and variance components of the covariance models separately while maintaining parsimony to ensure numerical stability and overall efficiency. In specifying a structure for $\boldsymbol{\Gamma}$, it will be important to assess whether a patterned or unpatterned model is most appropriate. If the images for each task are taken in the same imaging session, then a patterned structure like those in Table 2 suffices, where the $d_{jk}$ term represents the time between images. If they are taken at times far enough apart such that temporal correlations are weakened (and possibly negligible), then a completely unstructured matrix as defined below for $k$ tasks will be most applicable.

$$\boldsymbol{\Gamma} = \begin{pmatrix} 1 & \rho_{t_1,t_2} & \rho_{t_1,t_3} & \cdots & \rho_{t_1,t_k} \\ \rho_{t_2,t_1} & 1 & \rho_{t_2,t_3} & \cdots & \rho_{t_2,t_k} \\ \rho_{t_3,t_1} & \rho_{t_3,t_2} & 1 & \cdots & \rho_{t_3,t_k} \\ \vdots & \vdots & \vdots & \ddots & \vdots \\ \rho_{t_k,t_1} & \rho_{t_k,t_2} & \rho_{t_k,t_3} & \cdots & 1 \end{pmatrix}, \tag{15}$$

where $\rho_{t_i,t_j}$ is the correlation between tasks $i$ and $j$. While this framework is promising, suitability of a Kronecker product covariance structure for reasons detailed in [148] may be an issue. Its appropriateness can be assessed by the tests developed in [135, 147]. If found to be unsuitable, alternatives described in [148] can be explored. However, the methodological and computational benefits that the Kronecker approach affords may still make it preferable depending on the level of unsuitability [135, 147].

## 5. Discussion and future directions

The recent explosion of complex functional brain network analyses has led to a paradigm shift in the neuroscientific community and catalyzed the development of the methods discussed in this review. While much has been done in this area, the noted methodological gaps reflect the need for more analytical tools. The most pressing *statistical* needs reside in the areas of weighted network analysis (Section 2.4), propagating network estimation error to network metrics (Section 3.6), and multivariate modeling and inferential methods (Sections 4.2 and 4.3). Developing informative descriptive metrics and addressing computational issues due to dimensionality are two important problems requiring statistical input in weighted network analysis. Quantifying the propagation of error from network estimation to the metrics discussed in Section 3 remains an important and largely untapped area. As discussed in Section 4, the paucity of analytical methods for brain network analysis is most salient in network modeling and inference. This area most requires the fusing of novel statistical approaches with network-based functional neuroimage analysis. Many avenues of research exist within this domain. Developing a multivariate modeling framework that accounts for the complex dependence structure of networks and allows assessing the effects of multiple variables of interest and local network features on the overall network structure is paramount. Evaluating how dynamic changes in network connectivity relate to brain dysfunction will require methodological extensions of this framework. Further, these frameworks will necessitate assessments of robustness and goodness-of-fit for the methods developed within them. The nascent area of brain information flow dynamics, which aims to understand how network topology supports brain activity, also remains fertile ground for statistical inquiry [122].

Combining network-related information within or across subjects, particularly with regard to community structure analysis, also presents a ripe area for innovative methods development. Approaches to assess consistency of and compare community structure within or across subjects require more work given that intra- and inter-subject variability often leads to the number, size, and composition of modules varying widely. Additionally, constructing group-based "representative" networks, often used for group-based community structure comparisons [112, 120, 121, 123, 124], remains challenging. [118] used an ERGM framework to produce representative networks that captured important average topological properties and nodal distributions of those properties in a group of networks, but their approach fails to appropriately incorporate anatomical information and is computationally intensive.

Given the nascency of network-based functional neuroimaging research and the variety of scientific questions of interest, it is unlikely that an optimal analysis method exists. A multi-faceted suite of complementary approaches will likely be needed to move the field forward in each area: network construction, network description, and network modeling and inference. Much of the methodological work thus far has come from computer science and statistical physics. The field would benefit immensely from an influx of statistical expertise. The multidisciplinary nature of the field requires collaborative research teams comprised of scientists from a wide variety of disciplines. This approach not only utilizes the individual expertise of each group member, but also engenders a unique cross-fertilization of knowledge among these scientists focused on complex functional brain network analysis.

The future of statistical network-based neuroimaging remains promising as long as we take a conscientious approach to developing methods that appropriately account for data complexity. More generally, complexity-based neuroimaging analysis, which subsumes network-based analysis, represents a new paradigm aimed at quantifying the complex patterns inherent in physiological systems. This systems based approach represents the frontier in neuroscience, statistics, and the sciences more generally. Incorporating innovative statistical methods within this paradigm will aid in revolutionizing our understanding of the human brain, psychiatric and neurological disorders such as depression and Parkinson's disease, and treatment for such disorders.

## Acknowledgements

This work was supported by NIBIB K25 EB012236-01A1 (Simpson), NINDS U18 NS082143-01 (Bowman), and Wake Forest Older Americans Independence Center (P30 21332) and the Sticht Center on Aging (Laurienti). The authors thank the editor, associate editor, and referees for their helpful comments that improved the paper. An earlier version of this manuscript can be found at arxiv.org (Simpson et al., arXiv:1302.5721 [stat.ME])

**References**

[1] Biswal, B.B., Mennes, M., Zuo, X., Gohel, S., Kelly, C., Smith, S.M., et al. Toward discovery science of human brain function. *Proc. Natl. Acad. Sci. USA*, 107: 4734-4739, 2010.

[2] Sporns, O. *Networks of the Brain*. The MIT Press, 2010.

[3] Ogawa, S., Lee, T.M., Kay, A.R., and Tank D.W. Brain magnetic resonance imaging with contrast dependent on blood oxygenation. *Proc. Natl. Acad. Sci. USA*, 87: 9868-72, 1990.

[4] Biswal, B.B., Yetkin F.Z., Haughton, V.M., and Hyde, J.S. Functional connectivity in the motor cortex of resting human brain using echo-planar MRI. *Magn Reson Med.*, 34: 537-41, 1995.

[5] Friston, K.J. Functional and effective connectivity in neuroimaging: a synthesis. *Human Brain Mapping*, 2: 56-78, 1994.

[6] Bullmore, E., and Sporns, O. Complex brain networks: graph theoretical analysis of structural and functional systems. *Nature Reviews Neuroscience*, 10: 186-198, 2009.

[7] Telesford, Q.K., Simpson, S.L., Burdette, J.H., Hayasaka, S., Laurienti P.J. The brain as a complex system: using network science as a tool for understanding the brain. *Brain Connectivity*, 1(4): 295-308, 2011.

[8] Bassett, D.S., and Bullmore, E.T. Human brain networks in health and disease. *Curr. Opin. Neurol.*, 22: 340-347, 2009.

[9] Fornito, A., Zalesky, A., Pantelis, C., and Bullmore, E.T. Schizophrenia, neuroimaging, and connectomics. *NeuroImage*, 62: 2296-2314, 2012.

[10] Wu, T., Wang, L., Chen, Y., Zhao, C., Li, K., and Chan, P. Changes of functional connectivity of the motor network in the resting state in Parkinson's disease. *Neuroscience Letters*, 460: 6-10, 2009.

[11] Minoshima, S., Giordani, B.J., Berent, S., Frey, K.A., Foster, N.L., and Kuhl, D.E. Metabolic reduction in the posterior cingulate cortex in very early Alzheimer's disease. *Ann Neurol*, 42: 85-94, 1997.

[12] Burdette, J.H., Minoshima, S., Vander Borght, T., Tran, D.D., and Kuhl, D.E. Alzheimer disease: improved visual interpretation of PET images by using three-dimensional stereotaxic surface projections. *Radiology*, 198(3): 837-43, 1996.

[13] Hagmann, P., Cammoun, L., Gigandet, X., Meuli, R, Honey, C.J., Wedeen, V.J., and Sporns, O. Mapping the structural core of human cerebral cortex. *PLoS Biol*, 6: e159, 2008.

[14] Rubinov, M., and Sporns, O. Complex network measures of brain connectivity: uses and interpretations. *Neuroimage*, 52: 1059-1069, 2010.

[15] Fornito, A., Zalesky, A., and Bullmore, E.T. Network scaling effects in graph analytic studies of human resting-state fMRI data. *Front. Syst. Neurosci.*, 4: 22, 2010.

[16] Sporns, O., Tononi, G., Kotter, R. The human connectome: a structural description of the human brain. *PLoS ONE*, 2: e1049, 2005.

[17] Wang, J., Wang, L., Zang, Y., Yang, H., Tang, H., Gong, Q., et al. (2009). Parcellation-Dependent Small-World Brain Functional Networks: A Resting-State fMRI Study. *Human Brain Mapping*, 30(5): 1511-1523.

[18] Honey, C.J., Kotter, R., Breakspear, M., and Sporns, O. Network structure of cerebral cortex shapes functional connectivity on multiple time scales. *Proc. Natl. Acad. Sci. USA*, 104: 10240-10245, 2007.

[19] Tzourio-Mazoyer, N., Landeau, B., Papathanassiou, D., Crivello, F., Etard, O., Delcroix, N., Mazoyer, B., and Joliot, M. Automated anatomical labeling of activations in SPM using a macroscopic anatomical parcellation of the MNI MRI single-subject brain. *NeuroImage*, 15: 273–289, 2002.

[20] van den Heuvel, M.P., Stam, C.J., Boersma, M., and Hulshoff Pol, H.E. Small-world and scale-free organization of voxel-based resting-state functional connectivity in the human brain. *Neuroimage*, 43: 528-539, 2008.

[21] Hayasaka, S., and Laurienti, P.J. Comparison of characteristics between region- and voxel-based network analysis in resting-state fMRI. *NeuroImage*, 50: 499-508, 2010.

[22] Power, J.D., Cohen A.L., Nelson, S.M., Wig, G.S., Barnes, K.A., Church, J.A., et al., Functional network organization of the human brain. *Neuron*, 72: 665-78, 2011.

[23] Smith, S.M., Miller, K.L., Gholamreza, S., Webster, M., Beckmann, C.F., Nichols, T.E., et al. Network modelling methods for fMRI. *NeuroImage*, 54: 875-891, 2011.

[24] Bowman, F.D., Zhang, L., Derado, G., and Chen, S. Determining functional connectivity using fMRI data with diffusion-based anatomical weighting. *NeuroImage*, 62: 1769-1779, 2012.

[25] Varoquaux, G., Gramfort, A., Poline, J.B., and Thirion, B. Brain covariance selection: better individual functional connectivity models using population prior. *arXiv:1008.5071v4 [stat.ML]*, 2010.

[26] Fox, M.D., Snyder, A.Z., Vincent, J.L., Corbetta, M., Van Essen, D.C., and Raichle, M.E. The human brain is intrinsically organized into dynamic, anticorrelated functional networks. *Proc. Natl. Acad. Sci. USA*, 102: 9673-9678, 2005.

[27] Cecchi, G.A., Rish, I., Thyreau, B., Thirion, F.B., Plaze, M., Paillere-Martinot, M.L., Martelli, C., Martinot, J., and Poline, J.. Discriminative network models of schizophrenia. In: *Neural Information Processing Systems*, 2009.

[28] Moussa, M.N., Steen, M.R., Laurienti, P.J., and Hayasaka, S. Consistency of network modules in resting-state fMRI connectiome data. *PLoS ONE*, 7:e44428, 2010.

[29] Zalesky, A., Fornito, A., and Bullmore, E.T. On the use of correlation as a measure of network connectivity. *NeuroImage*, 60: 2096-2106, 2012.

[30] Hlinka, J., Palus, M., Vejmelka, M., Mantini, D., and Corbetta, M. Functional connectivity in resting-state fMRI: Is linear correlation sufficient? *NeuroImage*, 54: 2218-2225, 2011.

[31] Achard, S., Salvador, R., Whitcher, B., Suckling, J., Bullmore, E.. A resilient, low-frequency, small-world human brain functional network with highly connected association cortical hubs. *Journal of Neuroscience*, 26: 63-72, 2006.

[32] Marrelec, G., Krainik, A., Duffau, H., Pélégrini-Issac, M., Lehéricy, S., Doyon, J., and Benali, H. Partial correlation for functional brain interactivity investigation in functional MRI. *NeuroImage*, 32: 228–237, 2006.

[33] Hlinka, J., Hartman, D., and Palus M. Small-world topology of functional connectivity in randomly connected dynamical systems. *Chaos*, 22: 033107, 2012.

[34] Curtis, C.E., Sun, F.T., Miller, L.M., and D'Esposito, M. Coherence between fMRI time-series distinguishes two spatial working memory networks. *NeuroImage*, 26: 177-183, 2005.

[35] Muller, K., Lohmann, G., Bosch, V., and von Cramon, D.Y. On multivariate spectral analysis of fMRI time series. *NeuroImage*, 14: 347– 356, 2001.

[36] Chang, C., and Glover, G. Time-frequency dynamics of resting-state brain connectivity measured with fMRI. *Neuroimage*, 50, 81–98, 2010.

[37] Ombao, H., and Van Bellegem, S. Evolutionary coherence of nonstationary signals. *IEEE Transactions on Signal Processing*, 56: 2259-2266, 2008.

[38] Fiecas, M., and Ombao, H. The generalized shrinkage estimator for the analysis of functional connectivity of brain signals. *Annals of Applied Statistics*, 5: 1102-1125, 2011.

[39] Ma, S., Calhoun, V.D., Eichele, T., Du, W., and Adali, T. Modulations of functional connectivity in the healthy and schizophrenia groups during task and rest. *NeuroImage*, 62: 1694-1704, 2012.

[40] Salvador, R., Martinez, A., Pomarol-Clotet, E., Sarro, S., Suckling, J., and Bullmore, E. Frequency based mutual information measures between clusters of brain regions in functional magnetic resonance imaging. *NeuroImage*, 35: 83-88, 2007.

[41] Hartman, D., Hlinka, J., Palus, M., Mantini, D., and Corbetta, M. The role of nonlinearity in computing graph-theoretical properties of resting-state functional magnetic resonance imaging brain networks. *Chaos*, 21: 013119, 2011.

[42] Dauwels, J., Vialatte, F., Musha, T., and Cichocki, A. A comparative study of synchrony measures for the early diagnosis of Alzheimer's disease based on EEG. *NeuroImage*, 49: 668–693, 2010.

[43] Stam, C.J., and van Dijk, B.W. Synchronization likelihood: an unbiased measure of generalized synchronization in multivariate data sets. *Physica D*, 163: 236-251, 2002.

[44] Pereda, E., Quiroga, R.Q., and Bhattacharya, J. Causal influence: Nonlinear multivariate analysis of neurophysical signals. *Prog. Neurobiol.*, 77: 1–37, 2005.

[45] Netoff, I., Caroll, T.L., Pecora, L.M., and Schiff, S.J. Detecting coupling in the presence of noise and nonlinearity. In: Schelter, J., Winterhalder, W., Timmer. *Handbook of Time Series Analysis*. Wiley-B.W, 2006.

[46] Winterhalder, M., Schelter, B., Hesse, W., Schwab, K., Leistritz, L., Klan, D., et al. Comparison of linear signal procesing techniques to infer directed interactions in multivariate neural systems. *Signal Process*, 85: 2137–2160, 2005.

[47] Varoquaux, G., Gramfort, A., Poline, J.B., and Thirion, B. Markov models for fMRI correlation structure: Is brain functional connectivity small world, or decomposable into networks? *J. Physiol.*, *doi:10.1016/j.jphysparis.2012.01.001*, 2012.

[48] Cribben, I., Haraldsdottir, R., Atlas, L.Y., Wager, T.D., Lindquist, M.A. Dynamic connectivity regression: Determining state-related changes in brain connectivity. *NeuroImage*, 61: 907-920, 2012.

[49] Kolaczyk, E.D. *Statistical Analysis of Network Data: Methods and Models*. Springer, 2009.

[50] Rubinov, M., and Sporns O. Weight-converving characterization of complex functional brain networks. *NeuroImage*, 56: 2068-2079, 2011.

[51] van Wijk, B.C.M., Stam, C.J., and Daffertshofer, A. Comparing brain networks of different size and connectivity density using graph theory. *PLoS ONE*, 5: e13701, 2010.

[52] Stam, C.J., and Reijneveld, J.C. Graph theoretical analysis of complex networks in the brain. *Nonlinear Biomedical Physics*, 1: 3, 2007.

[53] Watts, D.J., and Strogatz, S.H. Collective dynamics of small-world networks. *Nature*, 393: 440-442, 1998.

[54] Stam, C.J. Functional connectivity patterns of human magnetoencephalographic recordings: a 'small-world' network? *Neurosci. Lett.*, 355: 25-28, 2004.

[55] Laurienti, P.J., Joyce, K.E., Telesford, Q.K., Burdette, J.H., and Hayasaka, S. Universal fractal scaling of self-organized networks. *Physica A*, 390: 3608-3613, 2011.

[56] Ginestet, C.E., and Simmons A. Statistical parametric network analysis of functional connectivity dynamics during a working memory task. *NeuroImage*, 55: 688-704, 2011.

[57] Kinnison, J., Padmala, S., Choi, J., Pessoa, L. Network analysis reveals increased integration during emotional and motivational processing. *The Journal of Neuroscience*, 32: 8361-8372, 2012.

[58] Ginestet, C.E., Nichols, T.E., Bullmore, E.T., and Simmons, A. Brain network analysis: Separating cost from topology using cost-integration. *PLoS ONE*, 6: e21570, 2011.

[59] Joyce, K.E., Laurienti, P.J., and Hayasaka, S. Complexity in a brain-inspired agent-based model. *Neural Networks*, 33: 275-290, 2012.

[60] Newman, M.E.J. The structure and function of complex networks. SIAM Rev., 45: 167-256, 2003.

[61] Latora, V., and Marchiori, M. Efficient behavior of small-world networks. *Phys. Rev. Lett.*, 87: 198701, 2001.

[62] Onnela, J.P., Saramaki, J., Kertesz, J., Kaski, K. Intensity and coherence of motifs in weighted complex networks. *Phy. Rev. E., Stat Nonlinear Soft Matter Phys.*, 71: 065103, 2005.

[63] Saramäki, J., Kivelä, M., Onnela, J.P., Kaski, K. and Kertész, J. Generalizations of the clustering coefficient to weighted complex networks. *Phy. Rev. E., Stat Nonlinear Soft Matter Phys.* 75: 027105, 2007.

[64] Fagiolo, G. Clustering in complex directed networks. *Phy. Rev. E., Stat Nonlinear Soft Matter Phys.*, 76: 026107, 2007.

[65] Sporns, O., and Honey, C.J. Small worlds inside big brains. *Proc. Natl. Acad. Sci. USA*, 103: 19219-19220, 2006.

[66] Maslov, S. and Sneppen, K. Specificity and stability in topology of protein networks. *Science* 296: 910-913, 2002.

[67] Humphries, M.D. and Gurney, K. Network 'small-world-ness': A quantitative method for determining canonical network equivalence. PLoS ONE 3: e0002051, 2008.

[68] Sporns, O. and Zwi, J. The small world of the cerebral cortex. *Neuroinformatics* 2: 145-162, 2004.

[69] Telesford, Q.K., Joyce, K.E., Hayasaka, S., Burdette, J.H., and Laurienti, P.J. The ubiquity of small-world networks. *Brain Connectivity*, 1: 367-375, 2011.

[70] Barabasi, A.L., and Albert, R. Emergence of scaling in random networks. *Science*, 286: 509-512, 1999.

[71] Rubinov, M., Sporns, O., Thivierge, J., Breakspear, M. Neurobiologically realistic determinants of self-organized criticality in networks of spiking neurons. *PLoS Computational Biology*, 7: e1002038, 2011.

[72] Virkar Y., and Clauset A. Power-law distributions in binned empirical data. *arXiv:1208.3524v1 [physics.data-an]*, 2012.

[73] Newman, M.E.J. Assortative mixing in networks. *Phys. Rev. Lett.*, 89: 2087011-2087014, 2002.

[74] Leung, C.C., and Chau, H.F. Weighted assortative and disassortative networks model. *Physica A*, 378: 591-602, 2007.

[75] Freeman, L. A set of measures of centrality based on betweenness. *Sociometry*, 40: 35-41, 1977.

[76] Freeman, L. Centrality in social networks: Conceptual clarification. *Soc. Networks*, 1: 215-239, 1979.

[77] Bonacich, P. Power and centrality: A family of measures. *Am. J. Sociol.*, 92: 1170-1182, 1987.

[78] Borgatti, S.P., and Everett, M.G. A Graph-theoretic perspective on centrality. *Soc. Networks*, 28: 466-484, 2006.

[79] He, Y., Chen, Z. and Evans, A. Structural insights into aberrant topological patterns of large-scale cortical networks in Alzheimer's disease. *J. Neurosci.*, 28: 4756-4766, 2008.

[80] Buckner, R.L., Sepulcre, J., Talukdar, T., Krienen, F.M., Liu, H., Hedden, T., et al. Cortical hubs revealed by intrinsic functional connectivity: Mapping, assessment of stability, and relation to Alzheimer's disease. *J. Neurosci.*, 29: 1860-1873, 2009.

[81] Lynall, M.-E., Bassett, D.S., Kerwin, R., McKenna, P.J., Kitzbichler, M., Muller, U., and Bullmore, E. Functional connectivity and brain networks in schizophrenia. *J. Neurosci.*, 30: 9477-9487, 2010.

[82] Lohmann, G., Margulies, D.S., Horstmann, A., Pleger, B., Lepsien, J., Goldhahn, D., et al. Eigenvector centrality mapping for analyzing connectivity patterns in fMRI data of the human brain. *PLoS ONE*, 5: e10232, 2010.

[83] Borgatti, S.P. Centrality and network flow. *Soc. Networks*, 27: 55-71, 2005.

[84] Südhof, T.C. The synaptic vesicle cycle. *Annu. Rev. Neurosci.*, 27: 509-547, 2004.

[85] Beggs, J.M. and Plenz, D. Neuronal avalanches in neocortical circuits. *J. Neurosci.*, 23: 11167-11177, 2003.

[86] Girvan, M. and Newman, M.E.J. Community structure in social and biological networks. *Proc. Natl. Acad. Sci. U S A*, 99: 7821-7826, 2002.

[87] Fortunato, S. Community detection in graphs. *Phys. Rep.*, 486: 75-174, 2010.

[88] Newman, M.E.J. and Girvan, M. Finding and evaluating community structure in networks. *Phys. Rev. E Stat. Nonlin. Soft Matter Phys.*, 69: 026113, 2004.

[89] Danon, L., Diaz-Guilera, A., Duch, J., and Arenas, A. Comparing community structure identification. *J. Stat. Mech.*, P09008, 2005.

[90] Newman, M.E.J. Modularity and community structure in networks. *Proc. Natl. Acad. Sci. U. S. A.*, 103: 8577–8582, 2006.

[91] Fortunato, S., and Barthélemy, M. Resolution limit in community detection. *Proc. Natl. Acad. Sci. U. S. A.*, 104: 36-41, 2007.

[92] Good, B.H., de Montjoye, Y.-A., and Clauset, A. The performance of modularity maximization in practical contexts. *Phys. Rev. E*, 81: 046106, 2010.

[93] Ruan, J. and Zhang, W. Identifying network communities with a high resolution. *Phys. Rev. E Stat. Nonlin. Soft Matter Phys.*, 77: 016104, 2008.

[94] Newman, M.E.J. Analysis of weighted networks. *Phys. Rev. E Stat. Nonlin. Soft Matter Phys.*, 70: 056131, 2004.

[95] Leicht, E.A., and Newman, M.E. Community structure in directed networks. *Phys. Rev. Lett.*, 100: 118703, 2008.

[96] Aldecoa, R., and Marín, I. Deciphering network community structure by Surprise. *PLoS ONE*, 6: e24195, 2011.

[97] Blondel, V.D., Guillaume, J.-L., Lambiotte, R., and Lefebvre, E. Fast unfolding of community hierarchies in large networks. *J. Stat. Mech.*, P10008, 2008.

[98] Wu, K., Taki, Y., Sato, K., Sassa, Y., Inoue, K., et al. The overlapping community structure of structural brain network in young healthy individuals. *PLoS ONE*, 6: e19608, 2011.

[99] Palla, G., Derenyi, I., Farkas, I. and Vicsek, T. Uncovering the overlapping community structure of complex networks in nature and society. *Nature*, 435: 814-818, 2005.

[100] Lancichinetti, A., Radicchi, F., Ramasco, J.J., and Fortunato, S. Finding statistically significant communities in networks. *PLoS ONE*, 6: e18961, 2011.

[101] Kovács, I.A., Palotai, R., Szalay, M.S., and Csermely, P. Community landscapes: An integrative approach to determine overlapping network module hierarchy, identify key nodes and predict network dynamics. *PLoS ONE*, 5: e12528, 2010.

[102] Steen, M., Hayasaka, S., Joyce, K.E., and Laurienti, P.J. Assessing the consistency of community structure in complex networks. *Phys. Rev. E Stat. Nonlin. Soft Matter Phys.*, 84: 016111, 2011.

[103] Mucha, P.J., Richardson, T., Macon, K., Porter, M.A., and Onnela, J.-P. Community structure in time-dependent, multiscale, and multiplex networks. *Science*, 328: 876-878, 2010.

[104] Bassett, D.S., Wymbs, N.F., Porter, M.A., Mucha, P.J., Carlson, J.M., and Grafton, S.T. Dynamic reconfiguration of human brain networks during learning. *Proc. Natl. Acad. Sci. U. S. A.*, 108: 7641-7646, 2011.

[105] Guimerà, R., and Amaral, L.A. Functional cartography of complex metabolic networks. *Nature*, 433: 895-900, 2005.

[106] Guimerà, R., Sales-Pardo, M., and Amaral, L.A. Classes of complex networks defined by role-to-role connectivity profiles. *Nat. Phys.*, 3: 63-69, 2007.

[107] Viles, W., Balachandran, P., and Kolaczyk, E. Uncertainty propagation from network inference to network characterization. *New England Statistics Symposium*, April 21, 2012.

[108] Bassett, D.S., Bullmore, E., Verchinski, B.A., Mattay, V.S., Weinberger, D.R., and Meyer-Lindenberg, A. Hierarchical organization of human cortical networks in health and schizophrenia. *J. Neurosci.*, 28: 9239-9248, 2008.

[109] Stam, C., Jones, B., Nolte, G., Breakspear, M., and Scheltens, P. Small-world networks and functional connectivity in Alzheimer's disease. *Cereb. Cortex.*, 17: 92-99, 2007.

[110] Zalesky, A., Fornito, A. and Bullmore, E.T. Network-based statistic: Identifying differences in brain networks. *NeuroImage*, 53: 1197-1207, 2010.

[111] Zalesky, A., Cocchi, L., Fornito, A., Murray, M.M., and Bullmore, E. Connectivity differences in brain networks. *NeuroImage*, 60: 1055-1062, 2012.

[112] Meunier, D., Achard, S., Morcom, A., and Bullmore, E. Age-related changes in modular organization of human brain functional networks. *NeuroImage*, 44: 715-723, 2009.

[113] Hipp, J.F., Engel, A.K., and Siegel, M. Oscillatory synchronization in large-scale cortical networks predicts perception. *Neuron*, 69: 387–396, 2011.

[114] Fornito, A., Yoon, J., Zalesky, A., Bullmore, E.T., and Carter, C.S. General and specific functional connectivity disturbances in first episode schizophrenia during cognitive control performance. *Biol. Psychiatry*, 70: 64–72, 2011.

[115] Zhang, J., Wang, J., Wu, Q., Kuang, W., Huang, X., He, Y., and Gong, Q. Disrupted brain connectivity networks in drug-naive, first-episode major depressive disorder. *Biol. Psychiatry*, 70: 334–342, 2011.

[116] Simpson, S.L., Hayasaka, S., and Laurienti, P.J. Exponential random graph modeling for complex brain networks. *PLoS ONE*, 6: e20039, 2011.

[117] Robins, G.L., Pattison, P.E., Kalish, Y., and Lusher, D. An introduction to exponential random graph (p*) models for social networks. *Social Networks*, 29: 173-191, 2007.

[118] Simpson, S.L., Moussa, M.N., and Laurienti, P.J. An exponential random graph modeling approach to creating group-based representative whole-brain connectivity networks. *NeuroImage*, 60: 1117-1126, 2012.

[119] Song, M., Liu, Y., Zhou, Y., Wang, K., Yu, C., and Jiang, T. Default network and intelligence difference. *Conf. Proc. IEEE Eng. Med. Bio.l Soc.*, 2212-2215, 2009.

[120] Valencia, M., Pastor, M.A., Fernández-Seara, M.A., Artieda, J., Martinerie, J., and Chavez, M. Complex modular structure of large-scale brain networks. *Chaos*, 19: 023119, 2009.

[121] Joyce, K.E., Laurienti, P.J., Burdette, J.H., and Hayasaka, S. A new measure of centrality for brain networks. *PLoS ONE*, 5: e12200, 2010.

[122] Jirsa, V.K., Sporns, O., Breakspear, M., Deco, G., and McIntosh, A.R. Towards the virtual brain: network modeling of the intact and the damaged brain. *Arch. Ital. Biol.*, 148: 189-205, 2010.

[123] Zuo, X., Ehmke, R., Mennes, M., Imperati, D., Castellanos, F.X., Sporns, O., and Milham, M.P. Network centrality in the human functional connectome. *Cerebral Cortex*, 22: 1862-1875, 2012.

[124] Gratton, C., Nomura, E.M., Perez, F., and D'Espsito, M. Focal brain lesions to critical locations cause widespread disruption of the modular organization of the brain. Journal of Cognitive Neuroscience, 24: 1275-1285, 2012.

[125] Handcock, M.S. *Statistical models for social networks: Inference and degeneracy*. Dynamic Social Network Modelling and Analysis: Workshop Summary and Papers. eds. R. Breiger, K. Carley, and P.E. Pattison. Washington, DC: National Academy Press: 229-240, 2002.

[126] Rinaldo, A., Fienberg, S.E., Zhou, Y. On the geometry of discrete exponential families with application to exponential random graph models. *Electronic Journal of Statistics*, 3: 446–484, 2009.

[127] Krivitsky, P.N. Exponential-family random graph models for valued networks. *Electronic Journal of Statistics*, 6: 1100-1128, 2012.

[128] Desmarais, B.A., and Cranmer, S.J. Statistical inference for valued-edge networks: The generalized exponential random graph model. *PLoS ONE* 7: e30136, 2012.

[129] Hoff, P.D. Multiplicative latent factor models for description and prediction of social networks. *Comput. Math Organ. Theory*, 15: 261-272, 2009.

[130] Hoff, P.D. Bilinear mixed-effects models for dyadic data. *Journal of the American Statistical Association*, 100: 286-295, 2005.

[131] Krivitsky, P.N., Handcock, M.S., Raftery, A.E., and Hoff, P.D. Representing degree distributions, clustering, and homophily in social networks with latent cluster random effects models. *Social Networks*, 31: 204-213, 2009.

[132] Albert, P.S., and Shen, J. Modelling longitudinal semicontinuous emesis volume data with serial correlation in an acupuncture clinical trial. *Journal of the Royal Statistical Society, Series C (Applied Statistics)*, 54: 707-720, 2005.

[133] Tooze, J.A., Grunwald, G.K., and Jones, R.H. Analysis of repeated measures data with clumping at zero. *Stat. Methods. Med. Res.*, 11: 341-355, 2002.

[134] Simpson, S.L., Edwards, L.J., Muller, K.E., Sen, P.K., and Styner, M.A. A linear exponent AR(1) family of correlation structures. *Statistics In Medicine*, 29: 1825-1838, 2010.

[135] Simpson, S.L. An adjusted likelihood ratio test for separability in unbalanced multivariate repeated measures data. *Statistical Methodology*, 7: 511-519, 2010.

[136] Handcock, M.S., Raftery, A.E., and Tantrum, J. Model-based clustering for social networks. *Journal of the Royal Statistical Society A*, 170: 301-354, 2007.

[137] Hoff, P.D., Raftery, A.E., and Handcock, M.S. Latent space approaches to social network analysis. *Journal of the American Statistical Association*, 97: 1090-1098, 2002.

[138] Moreno, S., Kirshner, S., Neville, J., and Vishwanathan, S.V.N. Tied Kronecker product graph models to capture variance in network populations. In: *Proc. 48th Annual Allerton Conf. on Communication, Control, and Computing*, 1137-1144, 2010.

[139] Fournel, A.P., Reynaud, E., Brammer, M.J., Simmons, A., and Ginestet, C.E. Group analysis of self-organizing maps based on functional MRI using restricted frechet means. *arXiv:1205.6158v2 [stat.AP]*, 2012.

[140] Snijders, T.A.B., van de Bunt, G.G., and Steglich, C.E.G. Introduction to stochastic actor-based models for network dynamics. *Social Networks*, 32:44-60, 2009.

[141] Snijders, T.A.B., Koskinen, J., Schweinberger, M. Maximum likelihood estimation for social dynamics. *The Annals of Applied Statistics*, 4: 567-588, 2010.

[142] Hanneke, S., Fu, W., and Xing, E.P. Discrete temporal models of social networks. *The Electronic Journal of Statistics*, 4: 585-605, 2010.

[143] Westveld, A.H., and Hoff, P.D. A mixed effects model for longitudinal relational and network data, with applications to international trade and conflict. *The Annals of Applied Statistics*, 5: 843-872, 2011.

[144] Desmarais, B.A., and Cranmer, S.J. Statistical mechanics of networks: estimation and uncertainty. *Physica A*, 391: 1865–1876, 2012.

[145] Galecki, A.T. General class of correlation structures for two or more repeated factors in longitudinal data analysis. *Communications In Statistics-Theory and Methods*, 23: 3105-3119, 1994.

[146] Naik, D.N., and Rao, S.S. Analysis of multivariate repeated measures data with a Kronecker product structured correlation matrix. *Journal of Applied Statistics*, 28: 91-105, 2001.

[147] Simpson, S.L., Edwards, L.J., Styner, M.A., and Muller, K.E. Kronecker product linear exponent AR(1) correlation structures and separability tests for multivariate repeated measures. *arXiv:1010.4471v2 [stat.AP]*, 2012.

[148] Fonseca, T.C.O., and Steel, M.F.J. A general class of nonseparable space-time covariance models. E*nvironmetrics*, 22: 224-242, 2011.

Table 1. Classification of centrality measures (reproduced from [7])

| Centrality measures | | |
|---|---|---|
| Radial | Closeness | *Closeness-like measures* |
| | | Centroid |
| | | Immediate effects centrality |
| | | Information |
| | Degree | *Degree-like measures* |
| | | Graph-theoretical power index |
| | | Leverage |
| | | K-path |
| | | Status |
| | | Total effects centrality |
| | | **Eigenvector** |
| | | Iterated Standing |
| | | Power |
| | | Prestige |
| Medial | Betweenness | *Betweenness-like measures* |
| | | Distance-weighted fragmentation |
| | | Flow betweenness |
| | | K-betweenness |
| | | Mediative effects centrality |
| | | Random-walk betweenness |
| | | Rush |

Table 2. Stationary correlation structures that are continuous functions of distance

| Structure | $(j,k)$th element[a], $j \neq k$ | Params |
|---|---|---|
| LEAR | $\rho^{d_{\min} + \delta[(d_{jk} - d_{\min})/(d_{\max} - d_{\min})]}$ | 2 |
| AR(1) | $\rho^{d_{jk}}$ | 1 |
| DE | $\rho^{d_{jk}^{\nu}}$ | 2 |
| GAR(1) | $\rho^{d_{jk}} \dfrac{\Gamma(d_{jk} + \delta) F(\delta, d_{jk} + \delta; d_{jk} + 1; \rho^2)}{\Gamma(\delta)\Gamma(d_{jk} + 1) F(\delta, \delta; 1; \rho^2)}$ | 2 |
| Exponential | $\exp(-d_{jk}/\phi)$ | 1 |
| Gaussian | $\exp(-d_{jk}^2/\phi^2)$ | 1 |
| Linear | $(1 - \phi d_{jk}) I(\phi d_{jk} \leq 1)$ | 1 |
| Matern | $[1/\Gamma(\nu)](d_{jk}/2\phi)^{\nu} 2 K_v(d_{jk}/\phi)$ | 2 |
| Spherical | $\left[1 - (3d_{jk}/2\phi) + \left(d_{jk}^3/2\phi^3\right)\right] I(d_{jk} \leq \phi)$ | 1 |

NOTE:

[a]$d_{jk}$ − distance between $j^{th}$ and $k^{th}$ measurement of $i^{th}$ subject.

$\Gamma(\,\cdot\,)$ − gamma function.

$F(\theta_1, \theta_2; \theta_3; \theta)$ − hypergeometric function.

$K_v(\,\cdot\,)$ − modified Bessel function of the second kind of (real) order $v \geq 0$.

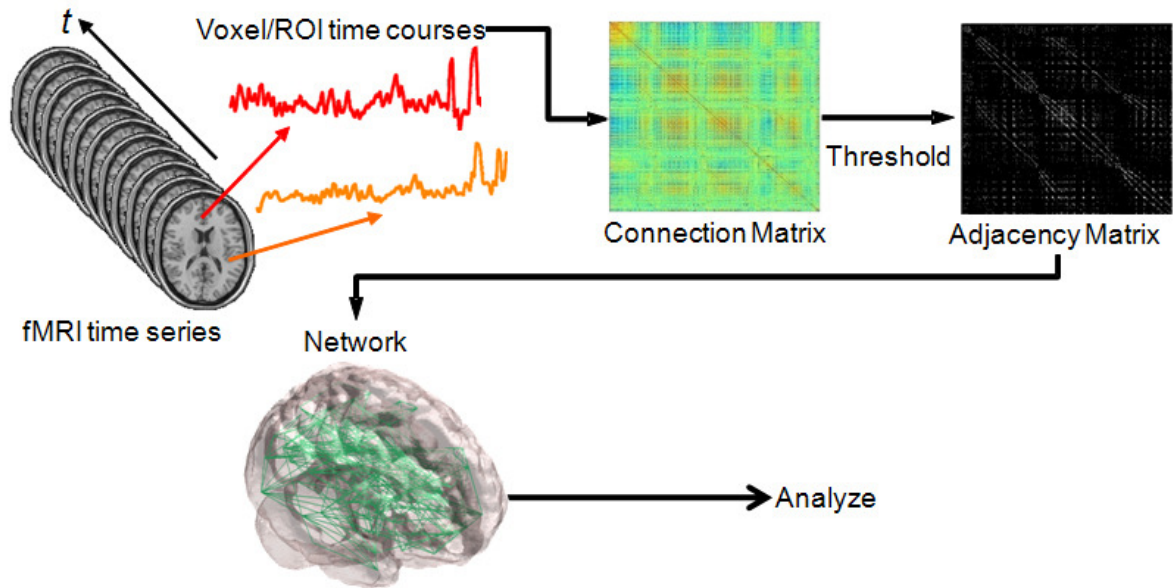

Fig 1. *Schematic for generating brain networks from fMRI time series data. Functional connectivity between brain areas is estimated based on time series pairs to produce a connection matrix. A threshold is commonly applied to the matrix to generate a binary adjacency matrix. From the adjacency matrix, various network analyses can be performed.*

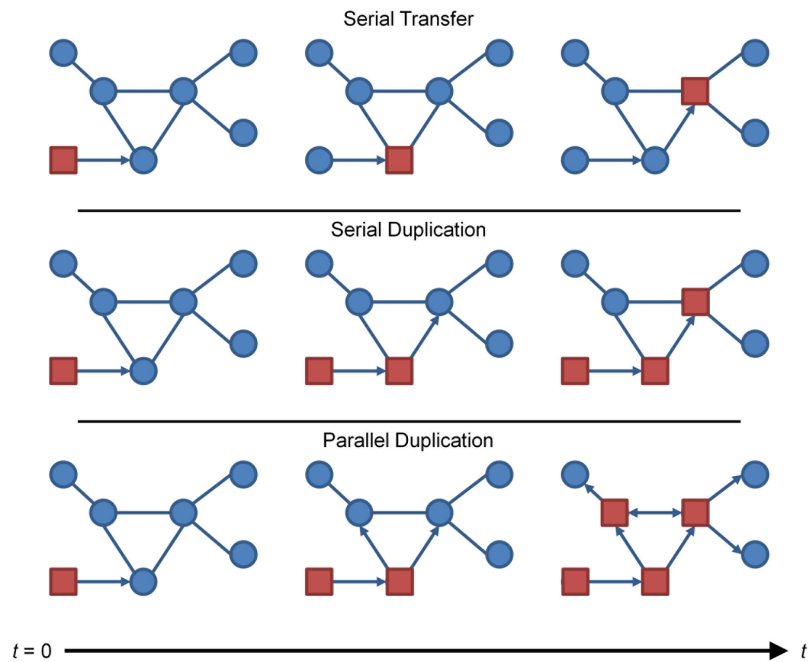

Fig 2. *Information flow patterns (reproduced from [7]). Squares indicate the existence of information at a node (circles). Arrows indicate the sequence and direction of information flow.*

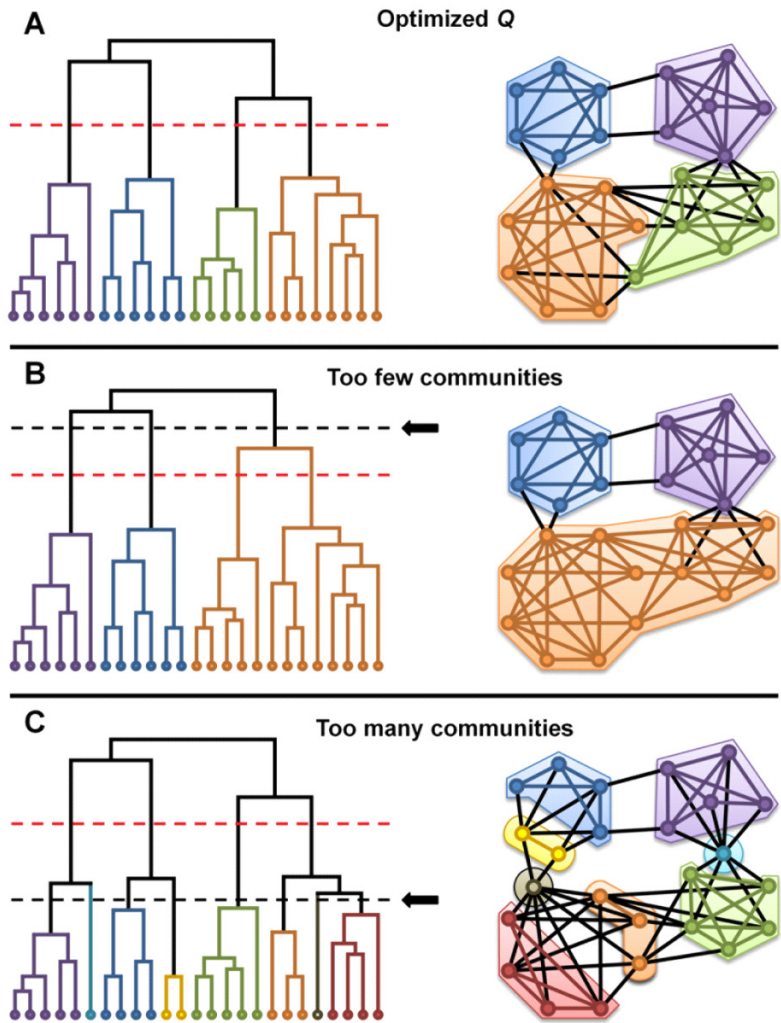

Fig 3. *Modularity analysis (reproduced from [7]). Depending on the level where the subdivisions are made (dashed line), the number of communities can change. (A) In this example network, the optimal Q yields four communities (indicated by the red dashed line). Shifting this line up or down (indicated by the dashed line with an arrow) produces a lower Q value that yields a suboptimal community structure. (B) As the line shifts higher, fewer communities are formed (approaching a single community comprising all nodes). (C) As the line shifts lower, more communities are formed (approaching every node being in their own community).*

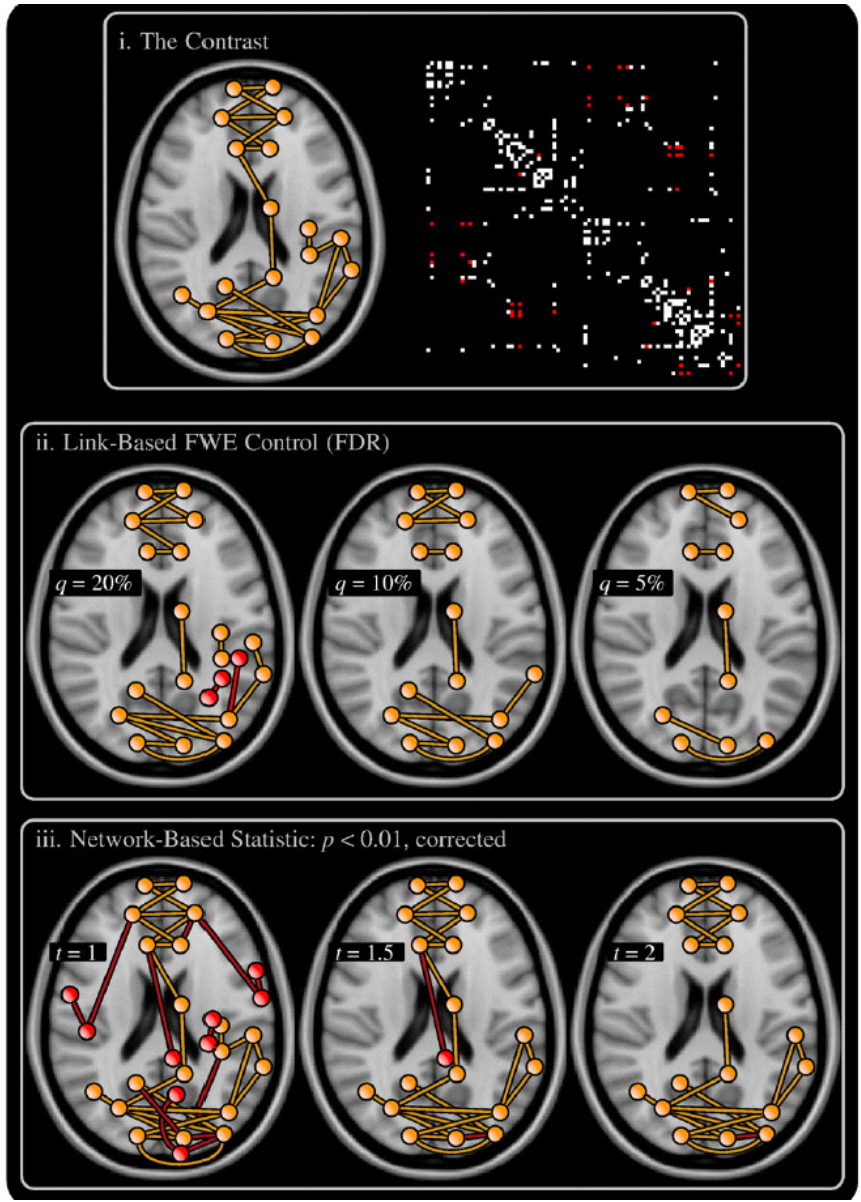

Fig 4 (reproduced from [110]). *The network-based statistic (NBS) as well as edge-based FWE control provided by the false discovery rate (FDR) were used to detect a contrast that was simulated between two groups: (i) a connected component, referred to as the contrast, was disrupted in one of the groups to yield a contrast-to-noise ratio of unity between the two groups. The red blocks of the adjacency matrix indicate edges comprising the contrast, while the white blocks indicate the other edges that were tested but were not part of the contrast. (ii) The FDR was used to identify the component using false discovery rate thresholds of $q = 5$, $10$ and $20\%$. (iii) The NBS was then used with primary (t-statistic) thresholds of $t = 1$, $1.5$ and $2$. True positives, colored orange, correspond to connections that were part of the contrast and correctly identified as such, while false positives, colored red, correspond to connections that were not part of the contrast but incorrectly identified as such. Each component identified by the NBS satisfied $p < 0.01$. With edge-based FWE control, the full extent of the contrast only became evident for a liberal false discovery rate threshold. The true and [false] positive rates for each threshold were: FDR: $q = 5\%$: $0.3[0]$; $q = 10\%$: $0.5[0]$; $q = 20\%$: $0.7[0.006]$; and NBS: $t = 1$: $1[0.08]$; $t = 1.5$: $0.9[0.01]$; $t = 2$: $0.9[0.006]$. Nodes are depicted at their two-dimensional center of mass. The two components evident for $t = 2$ were each of sufficient size to be declared significant in their own right.*

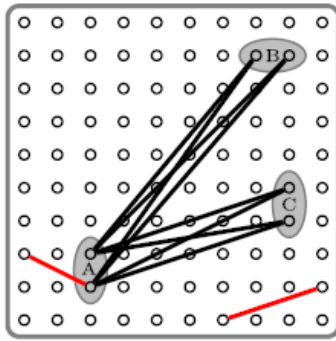

*Three interconnected regions:*
In this case, SPC identifies two separate pairwise clusters, A-B and A-C, whereas the NBS identifies one cluster corresponding to the entire network A-B-C. As such, the NBS only enables rejection of the null hypothesis at the level of the entire network A-B-C, whereas SPC enables rejection of the null hypothesis for A-B and A-C individually. The drawback of the NBS is that it detects the connection originating from region A, colored red, which can represent a false positive.

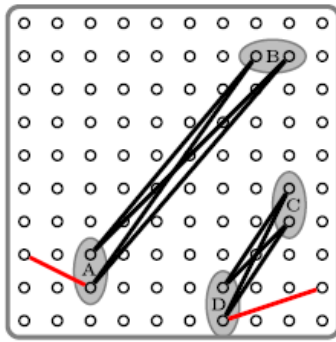

*Two pairs of interconnected regions:*
In this case, the NBS and SPC both identify two separate clusters corresponding to A-B and C-D. The null hypothesis can be rejected individually for A-B and A-C with both methods. The drawback of the NBS is that it detects the connections originating from regions A and D, colored red, which could be false positives. These false positives are disregarded by SPC because neither of them form *pairwise* neighbors.

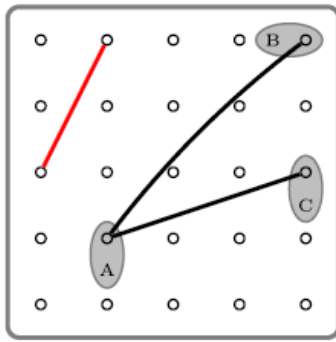

*Coarser sampling resolution:*
In this case, SPC does not identify any effect due to the lack of any pairwise relations; however, the NBS correctly identifies the cluster corresponding to the network A-B-C.

Fig 5 (reproduced from [111]). *Three examples illustrating the differences between the NBS and SPC. Circles represent nodes, while lines represent a supra-threshold connection. Black connections correspond to an experimental effect (true positives), while red connections correspond to false positives that survived the cluster-forming threshold.*